\documentstyle[12pt]{article}

\font\twlgot =eufm10 scaled \magstep1 \font\egtgot =eufm8
\font\sevgot =eufm7 \font\twlmsb =msbm10 scaled \magstep1
\font\egtmsb =msbm8 \font\sevmsb =msbm7

\newfam\gotfam
\def\pgot{\fam\gotfam\twlgot}
\textfont\gotfam\twlgot \scriptfont\gotfam\egtgot
\scriptscriptfont\gotfam\sevgot
\def\got{\protect\pgot}
\newfam\msbfam
\textfont\msbfam\twlmsb \scriptfont\msbfam\egtmsb
\scriptscriptfont\msbfam\sevmsb
\def\Bbb{\protect\pBbb}
\def\pBbb{\relax\ifmmode\expandafter\Bb\else\typeout{You cann't use
Bbb in text mode}\fi}
\def\Bb #1{{\fam\msbfam\relax#1}}

\def\thebibliography#1{\section*{References}\list
  {[\arabic{enumi}]}{\settowidth\labelwidth{#1}\leftmargin\labelwidth
    \advance\leftmargin\labelsep
    \usecounter{enumi}}
    \def\newblock{\hskip .11em plus .33em minus .07em}
    \sloppy\clubpenalty4000\widowpenalty4000
    \sfcode`\.=1000\relax}

\def\op#1{\mathop{\fam0 #1}\limits}

\newcommand{\id}{{\rm Id\,}}

\newcommand{\beq}{\begin{equation}}
\newcommand{\eeq}{\end{equation}}
\newcommand{\ben}{\begin{eqnarray}}
\newcommand{\een}{\end{eqnarray}}
\newcommand{\be}{\begin{eqnarray*}}
\newcommand{\ee}{\end{eqnarray*}}
\newcommand{\bea}{\begin{eqalph}}
\newcommand{\eea}{\end{eqalph}}

\newcommand{\cT}{{\cal T}}

\newcommand{\rL}{{\rm L}}

\newcommand{\al}{\alpha}

\newcommand{\bt}{\beta}

\newcommand{\La}{\Lambda}

\newcommand{\m}{\mu}

\newcommand{\g}{\gamma}
\newcommand{\G}{\Gamma}
\newcommand{\th}{\theta}

\newcommand{\vt}{\vartheta}

\newcommand{\di}{{\rm dim\,}}

\newcommand{\si}{\sigma}
\newcommand{\Si}{\Sigma}

\newcommand{\wt}{\widetilde}

\newcommand{\ol}{\overline}
\newcommand{\dr}{\partial}
\newcommand{\ar}{\op\longrightarrow}
\newcommand{\ot}{\otimes}

\newcounter{theorem}
\newcounter{remark}
\newcounter{proposition}
\newcounter{lemma}
\newcounter{corollary}
\newcounter{definition}

\def\theremark{\arabic{remark}}

\def\thedefinition{\arabic{definition}}

\newcommand{\mar}[1]{}

\begin{document}
\hbox{}

\begin{center}

{\large\bf GAUGE GRAVITATION THEORY FROM THE GEOMETRIC VIEWPOINT}
\bigskip

{\sc G.SARDANASHVILY}

{\it Department of Theoretical Physics, Moscow State University}

\end{center}

\bigskip
\bigskip
{\small

This is the Preface to the special issue of {\it International
Journal of Geometric Methods in Modern Physics} {\bf 3}, N.1
(2006) dedicated to the 50th anniversary of gauge gravitation
theory. It addresses the geometry underlying gauge gravitation
theories, their higher-dimensional, supergauge, and
non-commutative extensions.

}

\bigskip
\bigskip

At present, Yang--Mills gauge theory provides a universal
description of the fundamental electroweak and strong
interactions. Gauge gravitation theory from the very beginning
\cite{uti} aims to extend this description to gravity. It started
with gauge models of the Lorentz, Poincar\'e groups and general
covariant transformations [2--4]. At present, gauge gravity is
mainly supergravity, higher-dimensional, non-commutative and
quantum gravity [5--8]. This Preface can not pretend for any
comprehensive analysis of gauge gravitation theories, but
addresses only the geometry underlying them. We apologize in
advance for that the list of references is far from to be
complete.

The first gauge model of gravity was suggested by Utiyama
\cite{uti} in 1956 just two years after birth of the gauge theory
itself. Utiyama was first who generalized the original gauge model
of Yang and Mills for $SU(2)$ to an arbitrary symmetry Lie group
and, in particular, to the Lorentz group in order to describe
gravity. However, he met the problem of treating general covariant
transformations and a pseudo-Riemannian metric (a tetrad field)
which had no partner in Yang--Mills gauge theory.

In a general setting, fiber bundles provide the adequate geometric
formulation of classical field theory where classical fields on a
smooth manifold $X$ are represented by sections of fiber bundles
over $X$. In particular, Yang--Mills gauge theory is a gauge
theory of principal connections on a principal bundle $P\to X$ and
associated bundles with a structure Lie group $G$. There is the
canonical right action of $G$ on $P$ such that $P/G=X$. A typical
fiber of $P$ is the group space of $G$, and $G$ acts on it by left
multiplications. Being $G$-equivariant, principal connections on
$P$ (i.e., gauge potentials) are identified to global sections of
the quotient bundle $C=J^1P/G$, where $J^1P$ is the first order
jet manifold of sections of $P\to X$ \cite{book00}. Gauge
transformations in Yang--Mills gauge theory are vertical
automorphisms of a principal bundle $P\to X$ over $\id X$. Their
group VAut$(P)$ is isomorphic to the group of global sections of
the group bundle $P_{Ad}$ associated to $P$. A typical fiber of
$P_{Ad}$ is the group $G$ acting on itself by the adjoint
representation. Under certain conditions, one can define the
Sobolev completion of the group VAut$(P)$ to a Lie group. In
classical field theory, it suffices to consider local
one-parameter groups of local vertical automorphisms of $P$ whose
infinitesimal generators are $G$-invariant vertical vector fields
on $P$. These vector fields are identified to global sections of
the quotient bundle $VP/G\to X$, where $VP$ is the vertical
tangent bundle of $P\to X$. They form a Lie algebra $\cal G$ over
the ring $C^\infty(X)$ of smooth real functions on $X$. It fails
to be a Lie algebra of a Lie group, unless its Sobolev completion
exists. If $P\to X$ is a trivial bundle, $\cal G$ is a
$C^\infty(X)$-extension (a gauge extension in the physical
terminology) of the Lie algebra of $G$. Let
\mar{g6}\beq
Y=(P\times V)/G \to X\label{g6}
\eeq
be a fiber bundle associated to a principal bundle $P\to X$ whose
structure group $G$ acts on the typical fiber $V$ of $Y$ on the
left. Any automorphism of $P$ induces a bundle automorphism of
$Y$, and any principal connection on $P$ yields an associated
connection on $Y$. Such a connection is given by the $TY$-valued
form
\mar{g7}\beq
A=dx^\m\ot(\dr_\m +A^i_\m(x^\nu,y^j)\dr_i) \label{g7}
\eeq
with respect to bundle coordinates $(x^\m,y^i)$ on $Y$. If $Y$ is
a vector bundle, an associated connection (\ref{g7}) takes the
form
\mar{g8}\beq
A=dx^\m\ot(\dr_\m - A^r_\m(x^\nu) I_r{}^i{}_j y^j\dr_i),
\label{g8}
\eeq
where $I_r$ are generators of a representation of the Lie algebra
of a group $G$ in $V$. In Yang--Mills gauge theory on a principal
bundle $P$, sections of a $P$-associated vector bundle (\ref{g6})
describe matter fields. A generalization of the notion of an
associated bundle is the category of gauge-natural bundles
\cite{fat03,fat04}. In particular, any automorphism of $P$ yields
an automorphism of an associated gauge-natural bundle. The above
mentioned bundle of principal connections $C=J^1P/G$ exemplifies a
gauge-natural bundle.

Studying gauge gravitation theories, it seems reasonable  to
require that they must incorporate Einstein's General Relativity
and, in particular, admit general covariant transformations. Fiber
bundles possessing general covariant transformations belong to the
category of natural bundles \cite{kol}. Given a natural bundle
$T\to X$, there exists the canonical lift, called the natural
lift, of any diffeomorphism of its base $X$ to a bundle
automorphism of $T$ called a general covariant transformation.
Accordingly, any vector field $\tau$ on $X$ gives rise to a vector
field $\ol\tau$ on $T$ such that $\tau\mapsto\ol\tau$ is a
monomorphism of the Lie algebra $\cT(X)$ of vector field on $X$ to
that on $T$. One can think of the lift $\ol\tau$ as being an
infinitesimal generator of a local one-parameter group of local
general covariant transformations of $T$. Note that $\cT(X)$ fails
to be a Lie algebra of a Lie group. In contrast with the Lie
algebra $\cal G$ of infinitesimal gauge transformations in
Yang--Mills gauge theory, $\cT(X)$ is not a $C^\infty(X)$-algebra,
i.e., it is not a gauge extension of a finite-dimensional Lie
algebra.

Gauge gravitation theories thus are gauge theories on natural
bundles. The tangent $TX$, cotangent $T^*X$ and tensor bundles
over $X$ exemplify natural bundles. The associated principal
bundle with the structure group $GL(n,\Bbb R)$, $n=\di X$, is the
fiber bundle $LX$ of linear frames in the tangent spaces to $X$.
It is also a natural bundle. Moreover, any gauge-natural bundle
associated to $LX$ is a natural bundle. Principal connections on
$LX$ are linear connections on the tangent bundle $TX\to X$. If
$TX$ is endowed with holonomic bundle coordinates $(x^\m,\dot
x^\m)$ with respect to the holonomic frames $\{\dr_\m\}$, such a
connection is represent by the $TTX$-valued form
\mar{g1}\beq
\G=dx^\m\ot(\dr_\m + \G_\m{}^\al{}_\bt\dot x^\bt\dot\dr_\al),
\label{g1}
\eeq
where $\{\dr_\m, \dot\dr_\m=\dr/\dr\dot x^\m\}$ are holonomic
frames in the tangent bundle $TTX$ of $TX$. The bundle of linear
connections
\mar{g15}\beq
C_\G=J^1LX/GL(n,\Bbb R) \label{g15}
\eeq
is not associated to $LX$, but is a natural bundle, i.e., it
admits general covariant transformations \cite{book00}.

General covariant transformations are sufficient in order to
restart Einstein's General Relativity and, moreover, the
metric-affine gravitation theory \cite{tmf}. However, one also
considers the total group Aut$(LX)$ of automorphisms of the frame
bundle $LX$ over diffeomorphisms of its base $X$ \cite{heh}. Such
an automorphism is the composition of some general covariant
transformation and a vertical automorphism of $LX$, which is a
non-holonomic frame transformation. Subject to vertical
automorphisms, the tangent bundle $TX$ is provided with
non-holonomic frames $\{\vt_a\}$ and the corresponding bundle
coordinates $(x^\m,y^a)$. With respect to these coordinates, a
linear connection (\ref{g1}) on $TX$ reads
\mar{g2}\beq
\G=dx^\m\ot(\dr_\m + \G_\m{}^a{}_b(x^\nu) y^b\dr_a), \label{g2}
\eeq
where $\{\dr_\m,\dr_a\}$ are the holonomic frames in the tangent
bundle $TTX$ of $TX$. A problem is that the Hilbert--Einstein
Lagrangian is not invariant under non-holonomic frame
transformations. To overcome this difficulty, one can additionally
introduce an $n$-tuple of frame $\vt_a=\vt_a^\m\dr_\m$ (or coframe
$\vt^a=\vt^a_\m dx^\m$) fields, which is a section of the frame
bundle $LX$. This section is necessarily local, unless $LX$ is a
trivial bundle, i.e., $X$ is a parallelizable manifold.

Furthermore, the vector bundle $TX$ possesses a natural structure
of an affine bundle. The associated principal bundle with the
structure affine group $A(n,\Bbb R)$ is the bundle $AX$ of affine
frames in $TX$. There is the canonical bundle monomorphism $LX\to
AX$ corresponding to the zero section of the tangent bundle
$TX=AX/GL(n,\Bbb R)$. Principal connections on $AX$ are affine
connections on $TX$. Written with respect to an atlas with linear
transition functions, such a connection is a sum $K=\G + \Theta$
of a linear connection $\G$ on $TX$ and a $VTX$-valued (soldering)
form $\Theta$, i.e., a section of the tensor bundle $T^*X\ot VTX$,
where $VTX$ is the vertical tangent bundle of $TX\to X$
\cite{book00}. Given linear bundle coordinates $(x^\m,y^a)$ on
$TX$, it reads
\mar{g3}\beq
K=dx^\m\ot(\dr_\m + \G_\m{}^a{}_b(x^\nu) y^b\dr_a +
\Theta_\m^a(x^\nu)\dr_a). \label{g3}
\eeq
Due to the canonical isomorphism $VTX=TX\times TX$, the soldering
form $\Theta=\Theta_\m^a dx^\m\ot\dr_a$ yields the $TX$-valued
form $\Theta_X= \Theta_\m^a dx^\m\ot\vt_a$, and {\it vice versa}.
For instance, $K$ (\ref{g3}) is a Cartan connection if
$\Theta_X=\th=dx^\m\ot\dr_\m$ is the canonical tangent-valued form
on $X$ (i.e., the canonical global section $\th={\bf 1}$ of the
group bundle $LX_{Ad}\subset T^*X\ot TX$). Note that there are
different physical interpretations of the translation part
$\Theta$ of affine connections. In the gauge theory of
dislocations, a field $\Theta$ describes a distortion [14--16]. At
the same time, given a linear frame $\vt_a$, the decomposition
$\th=\vt^a\ot\vt_a$ motivates many authors to treat a coframe
$\vt^a$ as a translation gauge field (see \cite{heh,blag} and
references therein). A spinor representation of the Poincar\'e
group is called into play, too \cite{tiembl2}.

Yang--Mills gauge theory also deals with a Higgs field, besides
gauge potentials and matter fields. A Higgs field is responsible
for a symmetry breaking. Given a principal bundle $P\to X$ with a
structure Lie group $G$, let $H$ be a closed (consequently, Lie)
subgroup $H$ of $G$. Then we have the composite bundle
\mar{g9}\beq
P\ar^{\pi_{P\Si}} P/H\ar X, \label{g9}
\eeq
where $P\to P/H$ is a principal bundle with the structure group
$H$ and $P/H\to X$ is a $P$-associated bundle with the structure
group $G$ acting on its typical fiber $G/H$ on the left. In
classical gauge theory on $P\to X$, a symmetry breaking is defined
as a reduction of its structure group $G$ to the subgroup $H$ of
exact symmetries, i.e., $P$ contains an $H$-principal subbundle
called a $G$-structure [2,19--23]. There is one-to-one
correspondence $P^h=\pi^{-1}_{P\Si}(h(X))$ between the reduced
$H$-principal subbundles $P^h$ of $P$ and the global sections $h$
of the quotient bundle $P/H\to X$. These sections are treated as
classical Higgs fields. Any principal connection $A^h$ on a
reduced subbundle $P^h$ gives rise to a principal connection on
$P$ and yields an associated connection on $P/H\to X$ such that
the covariant differential $D_{A^h}h$ of $h$ vanishes. Conversely,
a principal connection $A$ on $P$ is projected onto $P^h$ iff
$D_Ah=0$. At the same time, if the Lie algebra ${\cal G}$ of $G$
is the direct sum
\mar{g13}\beq
{\cal G} = {\cal H} \oplus {\got m} \label{g13}
\eeq
of the Lie algebra ${\cal H}$ of $H$ and a subspace ${\got
m}\subset {\cal G}$ such that $ad(g)({\got m})\subset {\got m}$,
$g\in H$, then the pull-back of the ${\cal H}$-valued component of
any principal connection on $P$ onto a reduced subbundle $P^h$ is
a principal connection on $P^h$. This is the case of so-called
reductive $G$-structure \cite{godina03}.

Let $Y\to P/H$ be a vector bundle associated to the $H$-principal
bundle $P\to P/H$. Then, sections of the composite bundle $Y\to
P/H\to X$ describe matter fields with the exact symmetry group $H$
in the presence of Higgs fields. A problem is that the typical
fiber of a fiber bundle $Y\to X$ fails to carry out  a
representation of the group $G$, unless $G\to G/H$ is a trivial
bundle. It follows that $Y\to X$ is not associated to $P$ and, it
does not admit a principal connection in general. If $G\to G/H$ is
a trivial bundle, there exists its global section whose values are
representatives of elements of $G/H$. In this case, the typical
fiber of $Y\to X$ is $V\times G/H$, and one can provide it with an
induced representation of $G$ \cite{col68,mack}. Of course, this
representation is not canonical, unless $V$ itself admits a
representation of $G$. If $H$ is a Cartan subgroup of $G$, the
so-called non-linear realization of $G$ in a neighborhood of its
unit \cite{col,jos} exemplifies an induced representation.

In order to introduce a covariant differential on $Y\to X$, one
can use a principal connection on $Y\to P/H$
\cite{book00,sard92,ijgmmp06}.

Riemannian and pseudo-Riemannian metrics on a manifold $X$
exemplify classical Higgs fields. Let $X$ be an oriented
four-dimensional smooth manifold. The structure group
$GL_4=GL^+(4,\Bbb R)$ of $LX$ is always reducible to its maximal
compact subgroup $SO(4)$. The corresponding global sections of the
quotient bundle $LX/SO(4)$ are Riemannian metrics on $X$.
Accordingly, pseudo-Riemannian metrics of signature $(+,---)$ on
$X$ are global sections of the quotient bundle
\mar{b3203}\beq
\Si_{\rm PR}= LX/SO(1,3)\to X, \label{b3203}
\eeq
corresponding to the reduction of the structure group $GL_4$ of
$LX$ to its Lorentz subgroup $SO(1,3)$  \cite{iva,tra,sard80}.
Such a reduction need not exist, unless $X$ satisfies certain
topological conditions. From the physical viewpoint, the existence
of a Lorentz reduced structure comes from the geometric
equivalence principle and the existence of Dirac's fermion matter
\cite{iva,sardz}. The quotient bundle $\Si_{\rm PR}$ (\ref{b3203})
is associated to $LX$. It is a natural bundle. Its global section
$h$, called a tetrad field, defines a principal Lorentz subbundle
$L^hX$ of $LX$. Therefore, $h$ can be represented by a family of
local sections $\{h_a\}_\iota$ of $LX$ on trivialization domains
$U_\iota$ which take values in $L^hX$ and possess Lorentz
transition functions. One calls $\{h_a\}$ the tetrad functions,
Lorentz frames, or vielbeins. They define an atlas
$\Psi^h=\{(\{h_a\}_\iota, U_\iota)\}$ of $LX$ and associated
bundles with Lorentz transition functions. There is the canonical
imbedding of the bundle $\Si_{\rm PR}$ (\ref{b3203}) onto an open
subbundle of the tensor bundle $\op\vee^2T^*X$ such that its
global section $h=g$ is a pseudo-Riemannian metric
$g_{\m\nu}=h^a_\m h^b_\nu\eta_{ab}$ on $X$, which comes to the the
Minkowski metric $\eta$ with respect to an atlas $\Psi^h$. Since
the Lorentz group is a Cartan subgroup of $GL_4$, one can locally
put $g=\exp\{\si^{\al\bt}(x^\m)S_{\al\bt}\}(\eta)$, where
$S_{\al\bt}$ are non-Lorentz generators of $GL_4$, and treat the
parameter functions $\si^{\al\bt}(x^\m)$ as Goldstone fields
\cite{iva,ogi,ish}.

Any connection on a Lorentz principal bundle $L^hX$ is extended to
a connection on $LX$ and, thus, is a linear connection on $TX$ and
other associated bundles. It is called a Lorentz connection. The
covariant derivative of $g$ with respect to such a connection
vanishes. Thus, gauge theory on the linear frame bundle $LX$ whose
structure group is reduced to the Lorentz subgroup restarts the
metric-affine gravitation theory \cite{heh,tmf,blag,ant}.
Considering only Lorentz connections, we are in the case of
gravity with torsion [4,35--37]. If a connection is flat, its
torsion need not vanish. This is the case of a teleparallel
gravity [38--40] on a parallelizable manifold. Note that, since
orientable three-dimensional manifolds $M$ are parallelizable
\cite{sti}, any product $X=\Bbb R\times M$ is parallelizable (see
\cite{gue} for the case of a compact $X$).

Note that, if the structure group of $LX$ is reducible to the
Lorentz group $SO(1,3)$, it is also reducible to its maximal
compact subgroup $SO(3)$. The corresponding global section of the
quotient bundle $L^hX/SO(3)$ is a time-like unit vector field
$h_0$ on $X$ which provides a space-time decomposition of $TX$
\cite{sardz}. Moreover, there is the commutative diagram of
structure group reductions
\mar{g12}\beq
\begin{array}{ccc}
 GL_4 & \longrightarrow &  SO(1,3) \\
 \put(0,10){\vector(0,-1){20}} &
& \put(0,10){\vector(0,-1){20}}  \\
SO(4) & \longrightarrow &  SO(3)
\end{array} \label{g12}
\eeq
which leads to the well-known relation $g=2h_0\ot h_0 - g_R$
between pseudo-Riemannian $g$ and Riemannian $g_R$ metrics on $X$.

Any reduction of the structure group $GL_4$ of the linear frame
bundle $LX$ to the Lorentz one implies the corresponding reduction
of the structure group $A(4,\Bbb R)$ of the affine frame bundle
$AX$ to the Poincar\'e group. This is the case of the so-called
Poincar\'e gauge gravitation theory [3,4,43--47]. Since the
Poincar\'e group comes from the Wigner--In\"onii contraction of
the de Sitter groups $SO(2,3)$ and $SO(1,4)$ and it is a subgroup
of the conformal group, gauge theories on fiber bundles $Y\to X$
with these structure groups, reduced to the Lorentz one, are also
considered [45,48--52]. Because these fiber bundles fail to be
natural, the lift of the group $Diff(X)$ of diffeomorphisms of $X$
onto $Y$ should be defined \cite{lord,lord2}. One also meets a
problem of physical treating various Higgs fields. In a general
setting, one can study a gauge theory on a fiber bundle with the
typical fiber $\Bbb R^n$ and the topological structure group
$Diff(\Bbb R^n)$ or its subgroup of analytical diffeomorphisms
\cite{bor,kirsch}. Note that any paracompact smooth manifold
admits an analytic manifold structure inducing the original smooth
one \cite{nara}. The Poincar\'e gauge theory is also generalized
to the higher $s$-spin gauge theory of tensor coframes
$\vt_\m^{a_1\ldots a_{s-1}}dx^\m$ and generalized Lorentz
connections $A_\m^{a_1\ldots a_t,b_1\ldots b_t}$,
$t=1,\ldots,s-1$, which satisfy certain symmetry, skew symmetry
and traceless conditions \cite{vasil,vasil1}.

As was mentioned above, the existence of Dirac's spinor matter
implies the existence of a Lorentz reduced structure and,
consequently, a (tetrad) gravitational field. Note that, for the
purpose of gauge gravitation theory, it is convenient to describe
Dirac spinors in the Clifford algebra terms [60--62]. The Dirac
spinor structure on a four-dimensional manifold $X$ is defined as
a pair $(P^h, z_s)$ of a principal bundle $P^h\to X$ with the
structure spin group $L_s=SL(2,\Bbb C)$ and its bundle morphism
$z_s: P^h \to LX$ to the linear frame bundle $LX$ \cite{avis,law}.
Any such morphism factorizes
\mar{g10}\beq
P^h \to L^hX\to LX \label{g10}
\eeq
through some reduced principal subbundle $L^hX\subset LX$ with the
structure proper Lorentz group $L=SO^\uparrow(1,3)$, whose
universal two-fold covering is $L_s$. The corresponding quotient
bundle $\Si_{\rm T}=LX/L$ is a two-fold covering of the bundle
$\Si_{\rm PR}$ (\ref{b3203}). Its global sections are tetrad
fields $h$ represented by families of tetrad functions taking
values in the proper Lorentz group $L$. Thus, any Dirac spinor
structure is associated to a Lorentz reduced structure, but the
converse need not be true. There is the well-known topological
obstruction to the existence of a Dirac spinor structure
\cite{ger,wist}. For instance, a Dirac spinor structure on a
non-compact manifold $X$ exists iff $X$ is parallelizable.

Point out that the Dirac spinor structure (\ref{g10}) together
with the structure group reduction diagram (\ref{g12}) provides
the Ashtekar variables \cite{fat}.

Given a Dirac spinor structure (\ref{g10}), the associated Dirac
spinor bundle $S^h$ can be seen as a subbundle of the bundle of
Clifford algebras generated by the Lorentz frames $\{t_a\}\in
L^hX$ \cite{law,benn} (see also \cite{mosn}). This fact enables
one to define the Clifford representation
\mar{g11}\beq
\g_h(dx^\m)=h^\m_a\g^a \label{g11}
\eeq
of coframes $dx^\m$ in the cotangent bundle $T^*X$ by Dirac's
matrices, and introduce the Dirac operator on $S^h$ with respect
to a principal connection on $P^h$. Then, sections of a spinor
bundle $S^h$ describe Dirac spinor fields in the presence of a
tetrad field $h$. Note that there is one-to-one correspondence
between the principal connections on $P^h$ and those on the
Lorentz frame bundle $L^hX=z_s(P^h)$. Moreover, since the Lie
algebras of $G=GL_4$ and $H=L$ obey the decomposition (\ref{g13}),
any principal connection $\G$ on $LX$ yields a spinor connection
$\G_s$ on $P^h$ and $S^h$ \cite{pon,sard98a}. At the same time,
the representations (\ref{g11}) for different tetrad fields fail
to be equivalent. Therefore, one meets a problem of describing
Dirac spinor fields in the presence of different tetrad fields and
under general covariant transformations.

Due to the decomposition (\ref{g13}), there is also the canonical
lift of any vector field on $X$ onto the bundles $P^h$ and $S^h$
though they are not natural bundles [72-74]. However, this lift,
called the Kosmann's Lie derivative, fails to be an infinitesimal
generator of general covariant transformations. In order to solve
this problem, one can call into play the universal two-fold
covering $\wt{GL}_4$ of the group $GL_4$ which obeys the
commutative diagram
\mar{b3243}\beq
\begin{array}{ccc}
 \wt{GL}_4 & \longrightarrow &  GL_4 \\
 \put(0,-10){\vector(0,1){20}} &
& \put(0,-10){\vector(0,1){20}}  \\
L_s & \ar &  L
\end{array} \label{b3243}
\eeq
Let us consider the $\wt{GL}_4$-principal bundle $\wt{LX}\to X$
which is the two-fold covering bundle  of the frame bundle $LX$
{\cite{perc,law,dabr,swit}. This covering bundle is unique if $X$
is parallelizable, and it inherits general covariant
transformations of the linear frame bundle $LX$. However, spinor
representations of the group $\wt{GL}_4$ are infinite-dimensional.
Therefore, the $\wt{LX}$-associated spinor bundle describes
infinite-dimensional "world" spinor fields, but not the Dirac ones
\cite{heh,nee,sij}.

In a different way, we have the commutative diagram
\mar{b3222}\beq
\begin{array}{ccc}
 \wt{LX} & \ar^\zeta & LX \\
 \put(0,-10){\vector(0,1){20}} &
& \put(0,-10){\vector(0,1){20}}  \\
P^h & \ar & L^hX
\end{array} \label{b3222}
\eeq
for any Dirac spinor structure (\ref{g10}) \cite{sard98a,fulp}. As
a consequence, $\wt{LX}/\rL_s=LX/L=\Si_{\rm T}$. Let us consider
the $L_s$-principal bundle $\wt{LX}\to \Si_{\rm T}$ and the
associated spinor bundle $S\to \Si_{\rm T}$. It follows from the
diagram (\ref{b3222}) that, given a section $h$ of the tetrad
fiber bundle $\Si_{\rm T}\to X$, the pull-back of $S\to
 \Si_{\rm T}$ onto $h(X)\subset \Si_{\rm T}$ is exactly a spinor
bundle $S^h$ whose sections describe Dirac spinor fields in the
presence of a tetrad field $h$. Moreover, given the bundle of
linear connections $C_\G$ (\ref{g15}), the pull-back of the spinor
bundle $S\to \Si_{\rm T}$ onto $\Si_{\rm T}\times C_\G$ can be
provided with a connection and the Dirac operator $\cal D$
possessing the following property. Given a tetrad field $h$ and a
linear connection $\G$, the restriction of $\cal D$ to
$S^h=h(X)\times \G(X)$ is the familiar Dirac operator on the
spinor bundle $S^h\to X$ with respect to a spinor connection
$\G_s$ \cite{tmf,sard98a}. Thus, sections of the composite bundle
$S\to \Si_{\rm T}\to X$ describe Dirac spinor fields on $X$ in the
presence of different tetrad fields. If $X$ is parallelizable, one
can make $S\to X$ associated to the $\wt{GL}_4$-principal bundle
$\wt{LX}$, but not in a canonical way. Accordingly, $S\to X$
admits different lifts of vector fields on $X$. They differ from
each other in vertical fields on $S\to \Si_{\rm T}$ which are
infinitesimal generators of Lorentz gauge transformations.

Bearing in mind quantum field theory and unification models, one
considers spinor fields on Riemannian and pseudo-Riemannian
manifolds of any signature which possess additional symmetries.
Let us mention charged spinor fields on a Riemannian manifold $X$,
$\di X=n$. They are sections of a vector bundle associated to the
two-fold covering $P{{\rm Spin}^c}$ of the product $P_{SO}\times
P_{U(1)}$ of the $SO(n)$-principal bundle $P_{SO}$ of orthonormal
frames in $TX$ and a $U(1)$-principal bundle $P_{U(1)}$
\cite{law,mai,horie}. One says that $P{{\rm Spin}^c}$ defines a
spin$^c$-structure on $X$. It should be emphasized that a
spin$^c$-structure may prsent even if no spin structure exists.
For instance, any oriented four-dimensional Riemannian manifold
admits a spin$^c$-structure. If $X$ is compact, a
spin-(spin$^c$)-structure comes from Connes' commutative geometry
characterized by the spectral triple $({\cal A}, E,{\cal D})$ of
the algebra ${\cal A}=C^\infty(X,\Bbb C)$ of smooth complex
functions on $X$, the Hilbert space $E=L(X,S)$ of square
integrable sections of a spinor bundle $S\to X$, and the Dirac
operator $\cal D$ on $S$ \cite{conn,rennie}. This construction is
extended to non-compact manifolds $X$ \cite{ren}, globally
hyperbolic Lorentzian manifolds \cite{parf}, and pseudo-Riemannian
spectral triples \cite{stroh}. Non-abelian generalizations of a
spin$^c$-structure are also studied \cite{bala}.

In converse to that General Relativity can be derived from gauge
theory, the Kaluza--Klein theory generalized to non-Abelian
symmetries shows that higher-dimensional pseudo-Riemannian
geometry can lead to Yang--Mills gauge theory on fiber bundles
[88--90]. Namely, the following holds \cite{coqu,coqu98}. Let $E$
be a smooth manifold provided with a right action of a compact Lie
group $G$ such that all isotropy groups are isomorphic to a
standard one $H$. Then $E\to E/G$ is a fiber bundle with the
typical fiber $G/H$ and the structure group $N/H$, where $N$ is
the normalize of $H$ in $G$. Let $\g$ be a $G$-invariant metric on
$E$. Then it determines a $G$-invariant metric $\si$ on every
fiber of $E\to E/G$, a principal connection $A$ on this fiber
bundle and a metric $g$ on $E/G$, and {\it vice versa}. Moreover,
the scalar curvature of $\g$ falls into the sum of the scalar
curvature of $g$, the Yang--Mills Lagrangian for $A$ and the terms
depending on $\si$. This mathematical result however fails to
guarantee a perfect field model on a 'space-time' $E/G$. A problem
is to treat both the extra dimensions differing so markedly from
the space-time ones and the metric $\si$ whose components are
scalar fields on $E/G$ \cite{over}. It becomes a folklore that the
$D=11$ supergravity provides the most satisfactory solution of
this problem \cite{duff}.

Supergravity theory started with super-extensions of the (anti-)
De Sitter and Poincar\'e Lie algebras [5,94--96]. It is greatly
motivated both by the field-unification program and contemporary
string and brane theories [7,97--101]. There are various
superextensions of pseudo-orthogonal and Poincar\'e Lie algebras
[102--105]. Supergravity is mainly developed as their Yang--Mills
theory. The geometric formulation of supergravity as a partner of
gravity \cite{delig,binet} however meets difficulties.

First of all, it should be noted that the spin spectrum analysis
of frame and gauge fields fails to be correct, unless we are in
the case of a pseudo-Euclidean space which is not subject to
general covariant transformations. Furthermore, odd fields need
not be spinor fields. Ghosts and antifields in BRST field theory
exemplify odd fields. Among graded commutative algebras, the
Arens--Michel algebras of Grassmann type are most suitable for the
superanalysis \cite{bruz99}, but Grassmann algebras $\La$ of
finite rank are usually called into play. There are several
variants of supergeometry over such an algebra \cite{bart,bart93}.
They are graded manifolds, smooth $H^\infty$-, $G^\infty$-
$GH^\infty$-supermanifolds, $G$- and DeWitt supermanifolds. A
graded manifold is a pair of a smooth manifold $X$ and a sheaf of
Grassmann algebras on $X$. A smooth $GH^\infty$-supermanifold is a
graded local-ringed space $(M,\got S)$ which is locally isomorphic
to $(B^{n,m},\cal S)$ where $B^{n,m}=\La^n_0\oplus\La^m_1$ is a
supervector space and $\cal S$ is a sheaf of superfunctions on
$B^{n,m}$ taking their values in a Grassmann subalgebra
$\La'\subset\La$. One separates the following variants: (i)
$\La'=\La$ of $G^\infty$-supermanifolds introduced by A.Rogers
\cite{rog80}, (ii) ${\rm rank}\,\La -{\rm rank}\,\L'\geq m$ of
$GH^\infty$- supermanifolds, and (iii) $\La'=\Bbb R$ of
$H^\infty$-supermanifolds. The condition in item (ii) guarantees
that odd derivatives can be well defined. For instance, this is
not the case of $G^\infty$-supermanifolds, unless $m=0$. It is
essential that the underlying space $M$ of a smooth supermanifold
is provided with a structure of a smooth manifold of dimension
$2^{{\rm rank}\,\La-1}(n+m)$. At the same time, smooth
supermanifolds are effected to serious inconsistencies. For
instance, the sheaf of derivations of $G^\infty$-superfunctions is
not locally free, and spaces of values of
$GH^\infty$-superfunctions fail to be naturally isomorphic. The
$G$-supermanifolds are free of such inconsistences. They are
locally isomorphic to $(B^{n,m}, {\cal G}_{n,m})$, where the sheaf
${\cal G}_{n,m}$ of $G$-superfunctions is isomorphic to the tensor
product ${\cal S}\ot\La$, where $\cal S$ is a sheaf of
$H^\infty$-superfunctions. However, it may happen that
non-isomorphic $G$-manifolds can possess isomorphic underlying
smooth manifolds. Smooth supermanifolds and $G$-supermanifolds
become the DeWitt supermanifolds if they are provided with the
non-Hausdorff DeWitt topology. This is the coarsest topology such
that the body map $B^{n,m}\to \Bbb R^n$ is continuous \cite{dew}.
There is the well-known correspondence between DeWitt
supermanifolds and graded manifolds.

Lie supergroups, principal superbundles, associated supervector
bundles and principal superconnections are considered in the
category of $G$-supermanifolds in an appropriate way \cite{bart}.
Thus, they can provide an adequate mathematical language of
superextension of Yang--Mills gauge theory. Moreover, on may hope
that, since $G$-supermanifolds are also smooth manifolds, the
theorem of reduction of a structure group can be extended to
principal superbundles, and then Higgs superfields and
supermetrics on supermanifolds can be introduced as true geometric
partners of classical Higgs fields and gravity \cite{iva86}.
However, a problem is that even coordinates and variables on
supermanifolds are not real (or complex), but contain a nilpotent
summand. This is not a standard case of supergauge models
\cite{delig,binet}. Therefore, one should turn to graded manifolds
whose local bases consist of coordinates on a smooth body manifold
$X$ and odd generating elements of a Grassmann algebra $\La$. The
well-known Serre--Swan theorem extended to graded manifolds states
that, given a smooth manifold $X$, a Grassmann exterior algebra
generated by elements of a projective $C^\infty(X)$-module of
finite rank is isomorphic to the Grassmann algebra of graded
functions on a graded manifold with a body $X$ \cite{jmp05}. The
theory of graded principal bundles and connections has been
developed [116--117], but it involves Hopf algebras and looks
rather sophisticated. At the same time, graded manifolds provide
the adequate geometric formulation of Lagrangian BRST theory,
where supergauge transformations are replaced with a nilpotent
BRST operator \cite{barn,cmp04}.

There are strong reasons  to think that, due to quantum gravity,
space-time coordinates become non-commutative \cite{dopl}.
Non-commutative field theory is known to emerge from many
mathematical and physical models \cite{doug}. There are several
approaches to describing non-commutative gravity. One of them lies
in the framework of Connes' non-commutative geometry of spectral
triples [122--124]. In particular, a gravitational action can be
represented by the trace of a suitable function of the Dirac
operator \cite{conn97}. A gauge approach is based on the
Seiberg--Witten map replacing the original product of gauge,
vielbein  and metric fields with the star one \cite{chams1}.
Different variants of the $q$-deformation (quantum groups)
\cite{castel,cerc} and the Moyal-like (twist) deformation
[128--131] of space-time algebras are also considered. Note that
the deformation by the twist leads to complex geometry and, in
particular, to complexified gravity \cite{chams2}.

\bigskip
\bigskip
\noindent {\large \bf Acknowledgement}
\bigskip

I am grateful to Professors Juan Maldacena, M.A. Vasiliev, John R.
Klauder, Yuri N. Obukhov, and Ali H. Chamseddine for their
contribution to this issue of the Journal

\end{document}